\pdfoutput=1

\documentclass[11pt]{article}

\usepackage[final]{acl}

\usepackage{times}
\usepackage{latexsym}

\usepackage[T1]{fontenc}

\usepackage[utf8]{inputenc}

\usepackage{microtype}

\usepackage{inconsolata}

\usepackage{graphicx}

\title{CoCoST: Automatic Complex Code Generation \\
with Online Searching and Correctness Testing}

\makeatletter
\newcommand{\printfnsymbol}[1]{%
  \textsuperscript{\@fnsymbol{#1}}%
} 
\makeatother
\author{
Xinyi He\textsuperscript{\rm 1}\thanks{\indent The contributions by Xinyi He, Jiaru Zou and Yun Lin have been conducted and completed during their internships at Microsoft.}\hspace{0.5em}
Jiaru Zou\textsuperscript{\rm 2}\printfnsymbol{1}\hspace{0.5em}
Yun Lin\textsuperscript{\rm 3}\printfnsymbol{1}\hspace{0.5em}
Mengyu Zhou\textsuperscript{\rm 4}\thanks{\indent Corresponding author.}\hspace{0.5em} \\
\textbf{Shi Han} \textsuperscript{\rm 4}\hspace{0.5em}
\textbf{Zejian Yuan}\textsuperscript{\rm 1}\hspace{0.5em}
\textbf{Dongmei Zhang}\textsuperscript{\rm 4}\hspace{0.5em} \\
\textsuperscript{\rm 1} Xi'an Jiaotong University
\textsuperscript{\rm 2}  University of Illinois at Urbana-Champaign \\
 \textsuperscript{\rm 3} Peking University 
\textsuperscript{\rm 4} Microsoft Research\\
\texttt{\href{mailto:hxyhxy@stu.xjtu.edu.cn}{hxyhxy@stu.xjtu.edu.cn}},
\texttt{\href{mailto:linyun@stu.pku.edu.cn}{linyun@stu.pku.edu.cn}},
\texttt{\href{mailto:jiaruz2@illinois.edu}{jiaruz2@illinois.edu}},\\
\texttt{\href{yuan.ze.jian@xjtu.edu.cn}{yuan.ze.jian@xjtu.edu.cn}},
\texttt{\{\href{mailto:mezho@microsoft.com}{mezho}, \href{mailto:shihan@microsoft.com}{shihan}, \href{mailto:dongmeiz@microsoft.com}{dongmeiz}\}@microsoft.com}}

\usepackage{float}
\usepackage{subfig}
\usepackage{booktabs} 
\usepackage{amsmath}
\usepackage{amsthm}
\usepackage{amsfonts}       
\usepackage{nicefrac}       
\usepackage{xspace}
\usepackage{xcolor}
\usepackage{backnaur}
\usepackage{multirow}
\usepackage{makecell}
\usepackage{appendix}
\usepackage{enumitem}
\usepackage{circledsteps}
\usepackage{color}
\usepackage{colortbl}
\usepackage{graphicx}
\usepackage{amsmath}
\usepackage{caption}  
\usepackage{listings}
\usepackage{amssymb}

\theoremstyle{definition}

\newcommand{\refequ}[1]{Equation~(\ref{#1})}
\newcommand{\reffig}[1]{Figure~\ref{#1}}
\newcommand{\refsec}[1]{\S\ref{#1}} 
\newcommand{\reftab}[1]{Table~\ref{#1}}

\def\eg{\textit{e.g.}\xspace}

\begin{document}
\maketitle
\begin{abstract}
Large Language Models have revolutionized code generation ability by converting natural language descriptions into executable code. However, generating complex code within real-world scenarios remains challenging due to intricate structures, subtle bugs, understanding of advanced data types, and lack of supplementary contents. To address these challenges, we introduce the CoCoST framework, which enhances complex code generation by online searching for more information with planned queries and correctness testing for code refinement. Moreover, CoCoST serializes the complex inputs and outputs to improve comprehension and generates test cases to ensure the adaptability for real-world applications. 
CoCoST is validated through rigorous experiments on the DS-1000 and ClassEval datasets. Experimental results show that CoCoST substantially improves the quality of complex code generation, highlighting its potential to enhance the practicality of LLMs in generating complex code.

\end{abstract}

\section{Introduction}
\begin{figure*}[htbp]
    \centering
    \includegraphics[width=0.9\linewidth]{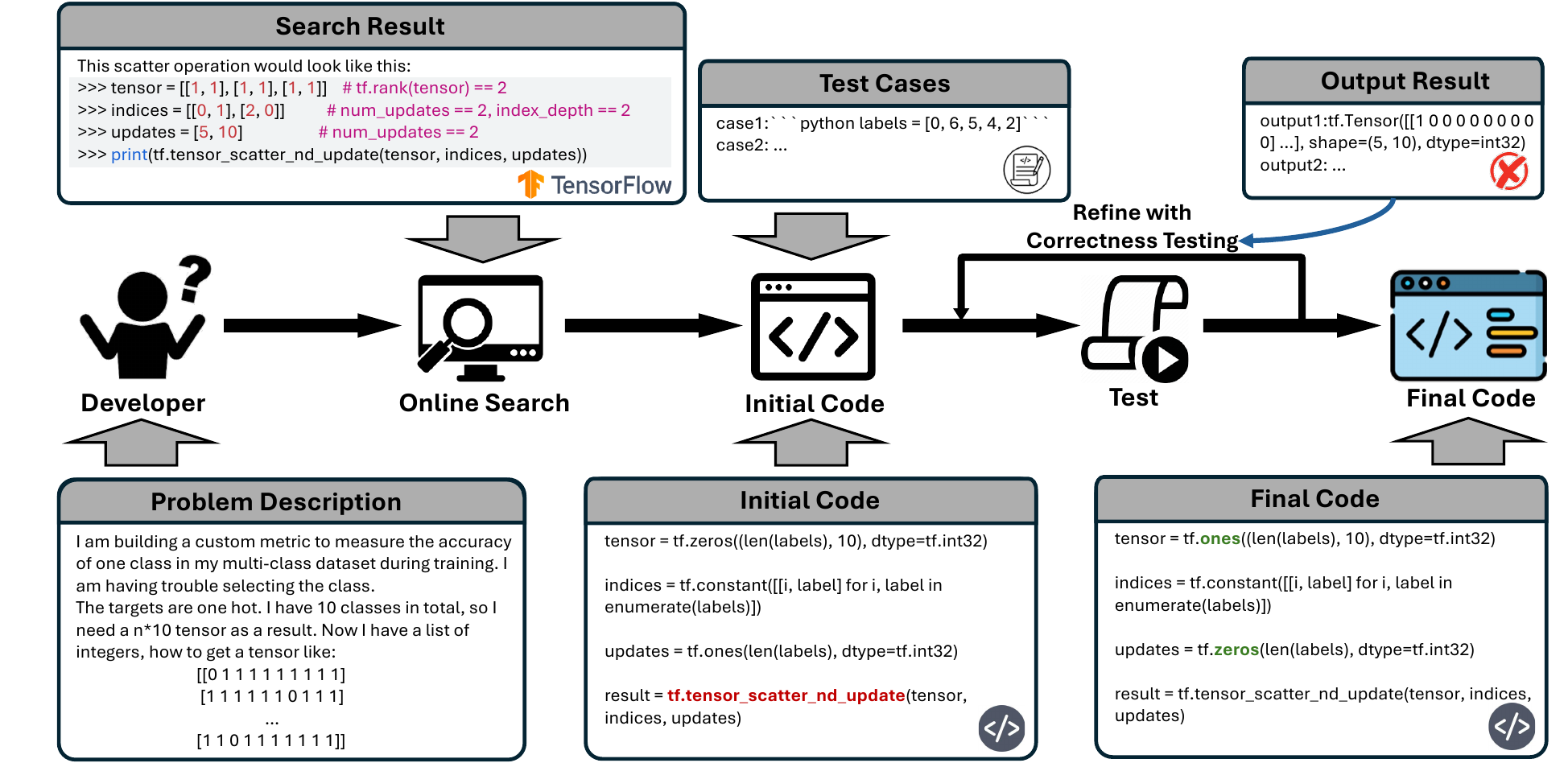}
    \caption{An Example of the Human Developer Code-writing Process Imitated by the CoCoST. After the problem is received, an online search is performed to simulate search results and create an initial version of the code. Test cases are then generated, and the code is executed to produce output results.  The code is refined based on the correctness of these results.}
    \label{fig:intro}
    \vspace{-3mm}
\end{figure*}

Automatic code generation from natural language descriptions is becoming more realistic, as large language models (LLMs) show their potential to generate accurate code~\cite{li2023starcoder, luo2023wizardcoder, rozière2024code}. 
Various methods have been proposed to improve the quality of LLM code generation, such as retrieving offline documents~\cite{zhou23docprompting,jiang2023selfevolve} and debugging generated code~\cite{zhang-etal-2023-self,chen2023teaching}.
However, complex code generation is a more difficult task, which involves intricate problem description, sophisticated code logic, and advanced data types \cite{Lai2022DS1000, du2023classeval, he2023text2analysis}. 
The existing methods struggle to address the arising challenges: 
    

\textit{Challenge 1}: 
 Offline documents cannot meet the diverse demands of code generation. In real-world scenarios, these demands often exceed the capabilities of limited offline documents. For example, problem descriptions may involve functions that are not covered by pre-collected documents. Additionally, complex code generation for diverse needs often entails highly complex logic and a series of transformation functions like the programming problem in \reffig{fig:intro}, where simple API examples in documents fail to provide adequate guidance.

\textit{Challenge 2}: In real-world situations, there is often a shortage of test cases (\eg, test cases in \reffig{fig:intro}) for automatic code generation. Most existing work depends heavily on pre-existing test cases in datasets~\cite{zhang-etal-2023-self, jiang2023selfevolve}, which are difficult to acquire directly in practical scenarios.

\textit{Challenge 3}: Hidden bugs in complex code require meticulous identification and refinement. Current techniques frequently enhance code by analyzing execution errors~\cite{zhang-etal-2023-self, jiang2023selfevolve}. But in the case of complex code, the executable code sometimes contains hidden bugs like the highlighted part of the initial code in \reffig{fig:intro}.


To address these challenges, we introduce a new code generation framework named \textbf{CoCoST}\footnote{The code will be open-sourced on https://github.com/microsoft/CoCoST.} (Automatic \textbf{Co}mplex \textbf{Co}de Generation with Online \textbf{S}earching and Correctness \textbf{T}esting) that improves the generation and refinement of complex code by LLMs through the planned online searching and automatic correctness testing steps. The intuition of CoCoST is straightforward: During the coding process, most human developers are not bothered by the above challenges, as illustrated in \reffig{fig:intro}. Developers can easily overcome these obstacles by searching online through engines (\eg, Google and Bing) for solutions, experiences, and guidelines. In addition, they can create test cases and execute code to ensure the correctness of the code logic.


To address \textit{Challenge 1}, CoCoST proposes an \textbf{online search} methodology. 
This process involves querying web search engines and then extracting pertinent information to construct LLM prompts. The approach presents several benefits:
(1) Retrieving information from the up-to-date blogs or Q\&A platforms, such as StackOverflow, facilitates the emulation of commonly used code patterns, thereby reducing the complexity of generated code. 
(2) Online search extends beyond the scope of static offline documentation, covering a wider range of problems without being confined to a predetermined set. Meanwhile, it reduces the effort developers need to expend in assembling documentation, thereby increasing the framework's level of automation.
Using problem descriptions as search queries can be difficult, because problems are generally intricate and include several components. Therefore, we propose an online search with \textbf{query generation through planning}.

To address \textit{Challenge 2}, we introduce \textbf{generation of test cases} during refinement. Several studies~\cite{codet, shinn2023reflexion} have attempted to generate tests. However, these methods often fall short when applied to the generation of complex code due to its intricate logic and outputs, which complicate the direct production of accurate tests (both the inputs and expected outputs for the solution code).
CoCoST utilizes LLMs to automatically generate test cases (the inputs for the code).
This strategy cleverly focuses on generating test cases without attempting to produce complete tests. It significantly simplifies the process of test case generation and facilitates its precise creation for complex code.

To address \textit{Challenge 3}, this work prioritizes \textbf{correctness testing} in refinement. 
During the refinement process, it is more critical to verify that the executed code produces the correct results rather than just checking the existence of the errors. CoCoST incorporates both the execution output results and the errors within the refinement prompts for LLMs to enhance the correctness. Moreover, during refinement, sophisticated data types and structures (within complex code itself, its inputs, and its execution results) are challenging for LLMs to understand, \eg, large Pandas DataFrames, and Matplotlib charts. Thus, CoCoST proposes \textbf{serialization of input and output} to convert them into understandable sequences before being processed by LLMs. Particularly those are excessively long or non-textual modalities.

We evaluated the effectiveness of CoCoST on two complex code generation datasets (DS-1000 and ClassEval). Compared with the existing state-of-the-art (SOTA) baseline, we achieve a 7.8\% improvement on DS-1000 and an average of 9.47\% on ClassEval.
Moreover, we analyze and discover that CoCoST requires models to have different capabilities such as planning, which vary according to the complexity of the problem. 
In summary, our main contributions are as follows.

\begin{itemize}
    \item We propose the novel CoCoST framework to generate complex code. CoCoST can be automatic in real-world scenarios.
    \item To generate complex code, we designed an online search method (query generation) in code generation for the first time to our knowledge.
    \item To refine hidden bugs in complex code, we prioritize correctness testing in refinement with test case generation and serialization of input and output data types.
    \item We conducted experiments on the DS-1000 and ClassEval datasets to demonstrate the effectiveness and universality of CoCoST.
\end{itemize}

\section{Related Work}

\textbf{Code generation datasets.} The realm of automated code generation has been propelled by benchmark datasets such as HumanEval~\cite{chen2021humaneval}, MBPP~\cite{austin2021mbpp}, and APPS~\cite{hendrycksapps2021}, which assess the proficiency of language models in generating executable code from descriptions. These datasets encompass a variety of programming problems, yet recent studies have sought to escalate the complexity of code generation tasks. Works like DS-1000~\cite{Lai2022DS1000}, ClassEval~\cite{du2023classeval} and Text2Analysis~\cite{he2023text2analysis} have introduced datasets targeting specialized domains, including data science, object-oriented class generation, and data analysis. These endeavors reflect an emerging trend towards enhancing models' abilities to produce sophisticated and domain-specific code structures. In this paper, we select datasets with complex code generation to evaluate CoCoST.

\noindent\paragraph{Retrieval-augmented code generation.} 
With the emergence of Large Language Models (LLMs), a variety of retrieval-augmented techniques have been developed to compensate for issues such as the inherent knowledge limitations. DocPrompt~\cite{zhou23docprompting} and SELVEVOLVE~\cite{jiang2023selfevolve} leverage document libraries or models as knowledge bases to improve code generation. However, their reliance on fixed document libraries limits the scope of information they can provide and confines the generated code to the context of these libraries. Furthermore, the prerequisite of preestablished document libraries prevents these approaches from being fully autonomous in real-world frameworks. Solutions such as WebGPT~\cite{nakano2022webgpt}, LaMDA~\cite{Thoppilan2022LaMDALM}, and FreshLLMs~\cite{vu2023freshllms} enhance the performance of natural language tasks by using online search or open web knowledge. However, because complex code generation often involves multiple steps and complexities, these methods struggle with direct application to complex code generation.

\noindent\paragraph{Code refinement.} 
Refine iteratively enhances generated code for greater precision. 
Self-Debug~\cite{chen2023teaching}, SELFEVOLVE~\cite{jiang2023selfevolve}, and Self-Edit~\cite{zhang-etal-2023-self} improve code generation by refining code through the resolution of errors identified during execution. These methods effectively address errors, while when it comes to complex code generation, subtle bugs also play a significant role in the overall error landscape. Moreover, relying on pre-existing tests from datasets in refinement limits their autonomy in real-world applications, where such tests may not be readily available.
CodeT~\cite{codet}, Reflexion~\cite{shinn2023reflexion}, and CODECHAIN~\cite{le2023codechain} seek to strengthen code generation by creating tests. But the tests they generate include not only the inputs for the solution code but also the expected outputs. This poses a substantial challenge for complex code generation, where the logic can be intricate and certain problems may not lend to straightforward ground truth generation.
\vspace{-2mm}
\begin{figure*}[htb]
    \centering
    \includegraphics[width=1\linewidth]{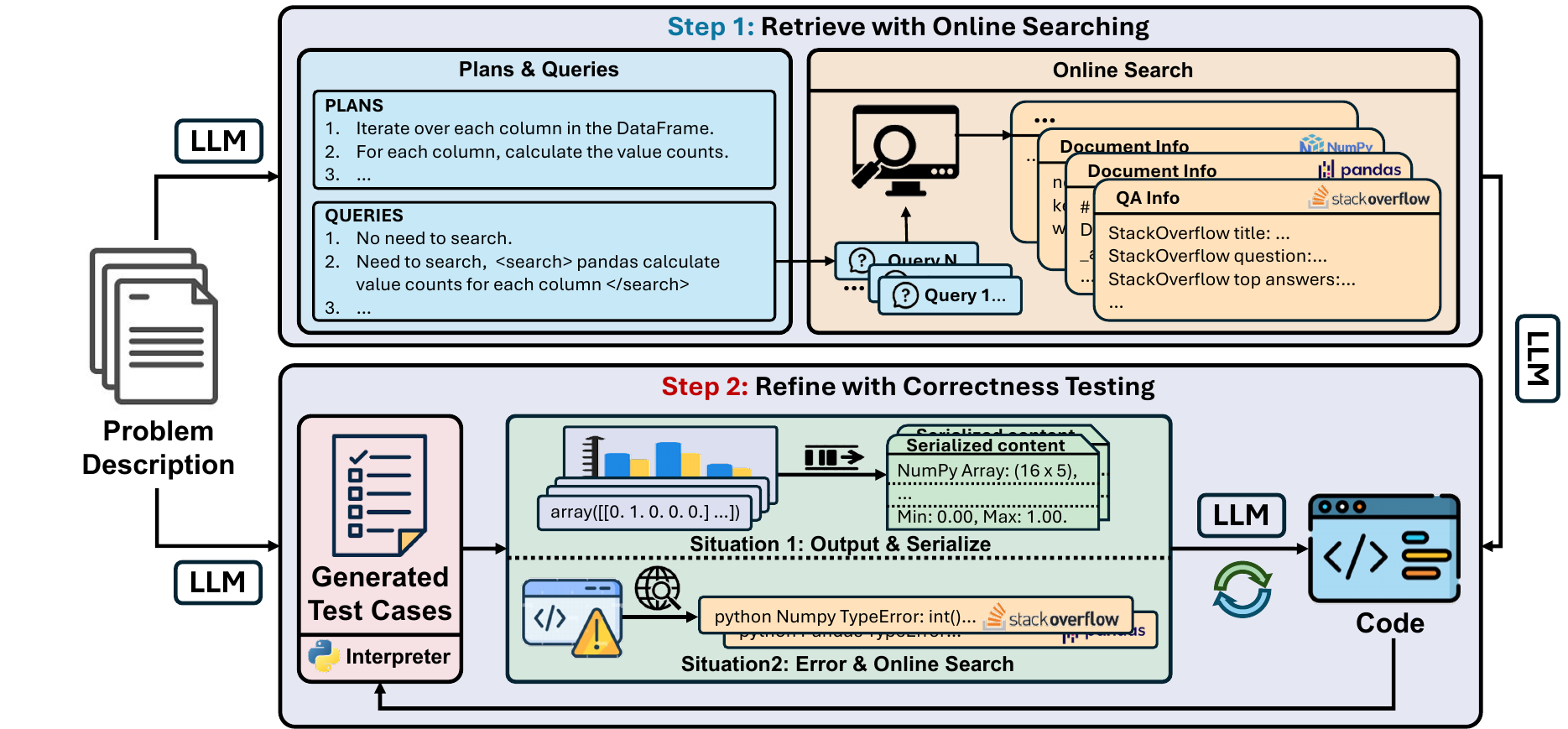}
    \caption{The Pipline of CoCoST. \textbf{Step 1:} LLM is employed to strategize the problem and formulate queries based on the outlined steps. These queries enable the retrieval of various information from the internet. A high-quality initial code can be obtained through effective planning and leveraging internet information. \textbf{Step 2:} LLM generates test cases to test the correctness of the initial code. The serialization of test results serves as crucial input for the subsequent cycle of code refinement. Through iterative refinement processes, the quality of the initial code can be significantly improved.}
    \label{fig:enter-label}
    \vspace{-4mm}
\end{figure*}

\section{Methodology}

The code generation task involves predicting a solution code $W$ given a problem description $D$. When given an input $i$, the execution of code $W$ produces an output result $o$ and a potential error $e$, where both $o$ and $e$ can be empty $\varnothing$. 
The generated codes are evaluated against a set of test cases and ground truth $\{(t_j, g_j)\}_{j=1}^J$. The correctness of the code $W$ is determined by verifying $o_j = g_j \land e_j = \varnothing$ when all $i_j=t_j, j \in \{1, \ldots , J\}$.

In this work, we adopt a two-step approach for code generation, mirroring the way humans write code. The first step is retrieval, where relevant information is obtained through an online search and utilized by LLMs to generate initial code. The second step is refinement, where the initial code is refined based on the execution results, leading to the generation of the final version of the code.

\subsection{Retrieval}
The difficulty in achieving effective online retrieval lies in formulating optimal search queries. On the one hand, for complex code generation, the problems are intricate and may involve multiple challenges. Directly searching for solutions to such problems is inaccurate and difficult. On the other hand, it is challenging that match queries directly through methods for offline documents like similarity calculations, due to the nature of online libraries. So we propose generating queries through planning to solve the challenge.

The retrieval process is divided into three steps: 1. Search queries $Q = \{q_1, \ldots, q_{N}\}$ are generated through planning. 2. Conducting online searches using these queries to obtain relevant background information $INFO = \{info_1, \ldots, info_{M}\}$. 3. The initial code $W_0$ is generated by the LLMs $\theta$ with the information obtained $INFO$:

{\small
\begin{equation}
    \widehat{W}_0 \sim p_\theta(.|D, INFO)
\end{equation}
}

\subsubsection{Generation Query through Planning}

To generate more targeted queries, we initiate the process by using LLMs to do planning regarding the given problem. The planning phase involves outlining the natural language steps $P = \{plan_1, \ldots, plan_N\}$ required to address the problem. 
Later, the assessment involves utilizing LLMs to determine whether each planning step requires an online search. Subsequently, the planning steps identified as necessitating online search are translated into queries $Q = \{q_1, \ldots, q_N\}$ for use in the subsequent search process.

{\small

\begin{equation}
    \widehat{P}, \widehat{Q} \ \sim p_\theta(.|D)
\end{equation}
}

\subsubsection{Online Search}

For the above-generated queries, we conduct an online search. In this study, we use the online search API\footnote{https://github.com/Nv7-GitHub/googlesearch} for the search process as \refequ{equ:search}. CoCoST can also be applied to private or domain-specific knowledge repositories as long as they are accessible via query, with details in \refsec{app:search}.

{
\small
\begin{equation}
    \{url_1, \ldots, url_{N_{u}}\} = search(q_j), j \in \{1, \ldots, N_{q}\}
\label{equ:search}
\end{equation}
}

where, $N_{q}$ is the number of queries for the problem, $N_{u}$ is the number of urls for one query. In this study, we use $N_{q}=1$ , $N_{u}=1$.

Through the analysis of the website distribution \reftab{tab:website}, we observed that more than 90\% of the URLs are concentrated on a total of 8 websites. Specific extraction rules are established for prominent websites such as StackOverflow to extract key information, facilitating a more comprehensive understanding of the website's content by subsequent models. Generic extraction rules are employed for extracting key information from other websites.
\vspace{-2mm}

{
\small
\begin{equation*}
info_{j,k} = extract(url_k), k \in \{1, \ldots, N_{u}\}
\end{equation*}
}

The information $INFO$ is composed of details from each query $q_j$, each URL $url_k$, with each piece of information $info_{j,k}$ extracted.

\subsection{Refinement}

Existing work \cite{chen2023teaching, jiang2023selfevolve} typically emphasizes the correctness of errors identified during the refinement process. However, we observe that refining code that produces error-free outputs is equally crucial during the refinement process. Therefore, we introduce correctness testing in \refsec{sec:correctness_testing}. Additionally, we propose methods for the generation of test cases and serialization of inputs and outputs during the refinement process.

\subsubsection{Correctness Testing}
\label{sec:correctness_testing}
Correctness testing refers to the refinement of generated code based on correctness, determined by analyzing errors and output results obtained during code execution. In the context of complex code generation, the intricate logic of the code makes it challenging for the LLMs to consider every detail during code generation, and precisely ascertain the results obtained at each step of the execution process. Consequently, some code may be executed without errors, producing output results that do not align with what is expected. 
Incorporating both the error and the output result into the refinement process allows the model to take advantage of self-correction mechanisms.

{\small
\begin{equation*}
\left\{
\begin{array}{ccl}
    e_{j, k}, o_{j, k} &=& execute(W_{j}, i_k), j \in \{1, \ldots, N_{f}\}\vspace{1ex}\\ 
    INFO_{e_{j, k}} &=& \{e_{j, k}, extract(search(e_{j, k}))\} \vspace{1ex}\\
    \widehat{W}_{j+1} &\sim & p_\theta(.|D, W_{j}, \{S_i, S_{o_j}, INFO_{e_j}\}_k),\\
    && \multicolumn{1}{r}{k\in \{1, \ldots, N_i\}}
\end{array}
\right.
\end{equation*}
}
where, $N_f$ is the total number of refinement steps, $N_i$ is the number of inputs. $i_K$ is the k-th input for the problem from \refequ{equ:test_case},  $S_i$ and $S_{o_j}$ is the serialization of input and output from \refequ{equ:serialize}.

\subsubsection{Generation of Test Cases}
Test cases are crucial, as they serve as indispensable inputs for the code execution in refinement. While, existing works in refining code predominantly rely on pre-existing test cases in datasets \cite{zhang-etal-2023-self, jiang2023selfevolve}, which are challenging to obtain directly in real-world scenarios. Moreover, some existing work~\cite{chen2023teaching} even uses the ground truth output of the test case to refine the code, which is even more challenging to obtain for complex code problems in real-world scenarios.  Because their problems involve various logical operations, deriving answers directly without code-based computations is demanding.

CoCoST introduces a generation of test cases with LLMs to adapt to real-world scenarios. 

{
\small
\begin{equation}
\left\{
    \label{equ:test_case}
    \begin{array}{ccl}
        \widehat{I} &\sim& p_\theta(.|D)\\
        I &=& \{i_1, \ldots, i_{N_i}\}
    \end{array}
\right.
\end{equation}
}

\subsubsection{Serialization of Input and Output}
Serialization of input and output makes them more intuitive and understandable for the model. For complex code, some inputs and outputs are intricate, such as Pandas DataFrames, PyTorch tensors, and Matplotlib PNG images. Understanding such inputs and outputs poses challenges for LLMs due to large matrices, image modalities, and so on. 

In this study, we serialize common data structures in Python as follows:
\begin{enumerate}[leftmargin=0pt,itemindent=\parindent, itemsep=2pt,topsep=0pt,parsep=0pt]
    \item For NumPy arrays, Pandas DataFrames, PyTorch tensors, and TensorFlow tensors, the serialization includes data truncated string, data type, data shape, and statistical information.
    \item For image structures (such as PNG images generated by the Matplotlib library), we serialize them into SVG (Scalable Vector Graphics) format for LLMs to comprehend.
\end{enumerate}

{
\small

\begin{equation}
    \label{equ:serialize}
    S_n = serialize(n), n \in \{i_k, o_{j,k}\}
\end{equation}

}

\begin{table*}[htb]
  \centering
  \caption{Main Results and Ablation Study for DS-1000. The base model for CoCoST is GPT-4. All metrics are represented as percentages. For each metric, the \textbf{bold} number indicates the highest performance.}
  \resizebox{0.75\textwidth}{!}{
    \begin{tabular}{lccccc}
    \toprule
    \multicolumn{1}{c}{\multirow{2}[4]{*}{Method}} & \multicolumn{4}{c}{Perturbation} & \multirow{2}[4]{*}{Total/Avg.} \\
\cmidrule{2-5}          & Origin & Surface & Semantic & Diff-Rewrite &  \\
    \midrule
    Codex & 44.93  & 37.94  & 34.35  & 16.94  & 39.20  \\
    DocPrompting & 53.95  & 50.00  & 38.39  & 21.05  & 43.30  \\
    Self-Debugging & 63.38  & 59.21  & 45.65  & 28.40  & 53.00  \\
    SELFEVOLVE & 66.23  & 67.11  & 48.70  & 33.95  & 57.10  \\
    Reflexion & 58.99 & 73.03 & 52.17 & 48.77 & 57.90 \\
    \midrule
    CoCoST & \textbf{71.71 } & 74.34  & \textbf{66.96 } & \textbf{53.09 } & \textbf{68.00 } \\
    \ \, w/o refinement of output & 68.42  & 69.74  & 62.61  & 48.77  & 64.10  \\
    \ \, w/o refinement of error & 68.20  & 73.03  & 62.61  & 49.38  & 64.60  \\
    \ \, w/o serialization & 70.18  & \textbf{75.00 } & 65.22  & 51.23  & 66.70  \\
    \ \, w/o generation of test case
    & 66.23  & 71.05  & 59.57  & 45.68  & 62.10  \\
    \ \, w/o online retrieval
    & 68.64  & 70.39  & 60.00  & 51.23  & 64.10  \\

    \ \, w/o all (GPT-4 only) & 64.47  & 69.74  & 56.96  & 43.83  & 60.20  \\

    \bottomrule
    \end{tabular}%
    }
  \label{tab:main_ds1000}%
\vspace{-3mm}
\end{table*}

\vspace{-3mm}

\section{Experiment}

\subsection{Experiment Setup}
\subsubsection{Datasets}
We conduct experiences on two complex code-generation datasets:

\textbf{DS-1000} \cite{Lai2022DS1000}: DS-1000 is a code generation benchmark with a thousand data science questions spanning seven Python libraries. The complexity of this dataset is manifested in two aspects. First, complexity arises from intricate logical reasoning required during code generation due to the complex nature of the problems. For example, on the DS-1000 dataset, the average length of problem descriptions is 140 words, whereas other commonly used code generation datasets such as HumanEval \cite{chen2021humaneval} and MBPP \cite{austin2021mbpp} have lengths of 23 and 15.7 words, respectively. Secondly, the input-output involves various complex data structures related to data science, making the code logic intricate during transformations of the data. Further details of DS-1000 implementation are shown in \refsec{app:ds-1000}.

\textbf{ClassEval} \cite{du2023classeval}: ClassEval is the first class-level Python code generation benchmark designed to evaluate code generation models' performance on a diverse set of object-oriented programming tasks. The dataset comprises a curated collection of 100 tasks. These tasks cover a wide range of concepts, including inheritance, polymorphism, encapsulation, etc. Each coding task is in the format of the class skeleton, outlining the target method description inside the class. The complexity of this dataset resides in its abstraction and hierarchical class structure. Tested models must generate large-scale code units and establish connections between each target method within the entire class, rather than focusing solely on individual functions. 

The dataset provides two prompt designs for LLMs with or without IF ability. In our experiments, we employ the class skeleton as the prompt for GPT-based models, a system prompt along with task instructions for the WizardCoder.


\begin{table*}[htb]
\small
  \centering
  \caption{Main Results and Ablation Study for ClassEval. All metric numbers are represented as percentages. For each metric, the \textbf{bold} number indicates the highest performance.}
  \resizebox{0.8\textwidth}{!}{
    \begin{tabular}{l|ccc|ccc}
    \toprule
    \multicolumn{1}{c|}{\multirow{2}[2]{*}{Method}} & \multicolumn{3}{c|}{Class-level} & \multicolumn{3}{c}{Method-level} \\
    \multicolumn{1}{c|}{} & Pass@1 & Pass@3 & Pass@5 & Pass@1 & Pass@3 & Pass@5 \\
    \midrule
    Instruct-StarCoder &    10.2  & 12.7   &  14.0    &  23.1    &   26.5   &   27.7\\
    SantaCoder & 8.6& 9.9& 10.0& 27.7& 33.0& 34.9\\
    Instruct-CodGen & 8.2& 12.3& 13.0& 24.9& 34.3& 37.1 \\
    WizardCoder &12.2& 20.0& 23.0& 35.2& 47.1& 51.1 \\
    Reflexion & 24.1 & 30.7 & 35.2 & 43.4 & 51.6 & 61.8 \\
    \midrule
    CoCoST &   \textbf{46.3}     &   \textbf{49.5}     &   \textbf{52.8}     &   \textbf{67.9}     &  \textbf{72.5}   &  \textbf{77.6}  \\ 
    \ \, w/o refinement of output &   43.5    &  46.8    &   51.4    &  66.4     &  69.0     & 73.4 \\
    \ \, w/o refinement of error &     46.2  &  \textbf{49.5}      &   51.7    &   \textbf{67.9}     &  \textbf{72.5}      &  77.2 \\
    \ \, w/o generation of test case & 42.7 & 47.9 &50.6 &65.9& 70.8 &72.4\\ 
    \ \, w/o online retrieval &    37.2   &  42.5     & 44.9  &  60.4     &  65.7     &  69.8 \\
    \ \, w/o all (GPT-4 only)  &  36.2     &  39.3     &  43.5     &   58.6   &  64.9     & 67.3 \\
    \bottomrule
    \end{tabular}%
    }
  \label{tab:main_classeval}%
\vspace{-3mm}
\end{table*}%
\subsubsection{Evaluation}
We employ the same evaluation methodology as the original datasets for both DS-1000 and ClassEval.

\textbf{DS-1000.} We follow the original dataset using Pass@1 accuracy. This evaluation is conducted across total and perturbations: Origin, Surface, Semantic, and Diff-Rewrite.

\textbf{ClassEval.} We follow the original dataset using Pass@$K$ metric. We calculate both class-leval and method-level Pass@$K$ with $K = 1, 3, 5$.

\subsubsection{Base LLMs}
This work primarily utilizes the GPT~\cite{openai2023gpt4} series as the LLM base model to validate the effectiveness of the framework. GPT-4 is utilized in \textit{gpt-4-32k-0613} version, while GPT-3.5 is utilized in the \textit{gpt-35-turbo-16k-0613} version. To further investigate the performance of CoCoST on both open-source and specialized code generation models, we have also employed WizardCoder~\cite{luo2023wizardcoder} as a base model with \textit{WizardCoder-Python-13B-V1.0} version.

\vspace{-1mm}

\subsubsection{Baselines}
For the DS-1000, we selected four LLM-based frameworks as baselines: DocPrompt~\cite{zhou23docprompting}, Self-Debugging~\cite{chen2023teaching}, SELFEVOLVE~\cite{jiang2023selfevolve} and Reflexion~\cite{shinn2023reflexion}.  
DocPrompting enhances the LLM by employing a fine-tuned retriever to fetch problem-relevant documentation from offline document pools. Self-debugging depends on a Python interpreter to instruct language models in revising Python code containing errors. SELFEVOLVE employs LLMs as both sources of knowledge and self-reflective programmers. Reflexion utilizes reflective feedback with generated tests and episodic memory to process task feedback. Details are shown in \refsec{app:baselines}.





For the ClassEval, we select five LLM-based code generation models and frameworks as baselines: Instruct-CodeGen\footnote{https://huggingface.co/sahil2801/instruct-codegen-16B}, SantaCoder~\cite{allal2023santacoder}, Instruct-StarCoder\footnote{https://huggingface.co/GeorgiaTechResearchInstitute/ starcoder-gpteacher-code-instruct }, WizardCoder~\cite{luo2023wizardcoder} and Reflexion~\cite{shinn2023reflexion}.

\subsection{Main Results}

Regarding the DS-1000 dataset, the main results are shown in \reftab{tab:main_ds1000}. CoCoST surpasses the current SOTA framework, SELFEVOLVE, by 10.9\%, establishing itself as the new SOTA. Especially under the Diff-Rewrite perturbation setting, CoCoST exceeds SELFEVOLVE by 19.95\%, which demonstrates the effectiveness of CoCoST in generating complex code. CoCoST employs online search and correctness testing to allow the model to imitate existing code patterns, thereby reducing the difficulty of generating new code and refining the details to further enhance the correctness of the code.

For the ClassEval dataset, the results are shown in \reftab{tab:main_classeval}. Our experiments demonstrate that CoCoST has an overall higher performance on both class-level and method-level Pass@$K$ evaluation. Specifically, CoCoST outperforms the Reflexion (best baseline model) significantly by an average of 19.5\% and 20.4\% on the Class and Method level.

\vspace{-1mm}

\subsection{Ablation Study}

\begin{table*}[htb]

  \centering
  \caption{Different Base Models Results for DS-1000 and ClassEval. All metric numbers are represented as percentages. For each metric in each section, the \textbf{bold} number indicates the highest performance.}
  \resizebox{0.8\textwidth}{!}{
    \begin{tabular}{l|cccc|c|cc}
    \toprule
    \multicolumn{1}{c|}{\multirow{2}[2]{*}{Method}} & \multicolumn{5}{c|}{DS-1000} & \multicolumn{2}{c}{ClassEval}\\
\cmidrule{2-8} 
& Origin & Surface & Semantic & Diff-Rewrite &   Total/Avg.& Class-level&Method-level\\
    \midrule
    GPT-4 & 64.47  & 69.74  & 56.96  & 43.83  & 60.20   & 43.5&67.3\\
    \ \, + retrieve & 66.23  & 71.05  & 59.57  & 45.68  & 62.10   & 50.6&72.4\\
    \ \, + refine &   68.64  & 70.39  & 60.00  & 51.23  & 64.10  & 44.9&69.8\\
    \ \, CoCoST & \textbf{71.71 } & \textbf{74.34 } & \textbf{66.96 } & \textbf{53.09 } & \textbf{68.00 }  & \textbf{52.8}&\textbf{77.6}\\
    \midrule
    GPT-3.5 & \textbf{57.02 } & 43.42  & 40.00  & 32.72  & 47.10   & 35.4&59.4\\
    \ \, + retrieve & 47.15  & 25.00  & 36.96  & 25.31  & 37.90   & 41.9&61.7\\
    \ \, + refine & 55.70  & \textbf{50.66 } & \textbf{44.35 } & \textbf{35.80 } & \textbf{49.10 }  & 42.8&62.3\\
    \ \, CoCoST & - & - & - & - & - & \textbf{45.8} & \textbf{64.7} \\
    \midrule
    WizardCoder &       \textbf{41.01}&       \textbf{21.71}&   \textbf{31.74}&       \textbf{16.05}&  \textbf{31.90} & \textbf{23.0}&\textbf{51.1}\\
    \ \, + retrieve &       15.79 &       9.21 &       12.17 &       9.88 &  13.00  & 18.2&41.8\\
    \ \, + refine &       39.69&       21.71&       30.00&       15.43&  30.80 & 22.3&50.7\\
    \bottomrule
    \end{tabular}%
    }
  \label{tab:base_model_ds1000}%
\vspace{-3mm}
\end{table*}%

In this work, to validate the effectiveness of CoCoST, we conduct different ablation studies, with results presented in \reftab{tab:main_ds1000} and \ref{tab:main_classeval}. Details on the ablation study are shown in \refsec{app:abla}.

\textbf{CoCoST significantly enhances the base model's ability to generate complex code.} Compared to the base model, CoCoST has shown improvements of 7.8\% on the DS-1000 dataset and an average of 9.47\% on ClassEval, demonstrating the effectiveness of the CoCoST. 

\textbf{Online search, generation of test cases, and serialization each contribute to the model's performance improvements.} Compared to CoCoST, after performing ablation studies, these features showed a decrease in performance of 3.9\%, 5.9\%, and 1.3\% respectively on the DS-1000 dataset.
The online search improves the model by providing common code patterns, which reduces the difficulty of the model in generating initial code. Serialization, by converting inputs and outputs into a sequential format, allows the model to more intuitively observe inputs and outputs that are too lengthy or are in non-textual modalities, thereby strengthening its ability to solve complex code problems.

\textbf{Online search outperforms offline retrieval in effectiveness and has a wider range of applicability.} As shown in \reftab{tab:main_ds1000}, using only online retrieval (the row w/o generation of the test case) outperforms DocPrompting, which is an offline retrieval approach. Moreover, in real-world scenarios, as opposed to specific datasets, the types of problems encountered are more diverse. The scalability of online retrieval enables them to effectively address a wide range of problems. However, offline retrieval systems struggle to encompass all relevant information comprehensively.

\textbf{During the refinement process, correctness testing is crucial, meaning that both the output result and error are equally important.} After separately conducting ablation studies on the output result and error, CoCoST shows a decrease of 3.9\% and 3.4\% respectively on the DS-1000 dataset, and an average of 2.7\% and 0.3\% on the ClassEval dataset. This indicates that the output result contributes more to the refinement process than the error. However, in previous works, the output result is often overlooked, which should not be the case, especially in the generation of complex code.
The evidence from the ablation study emphasizes the necessity of paying attention to the output results during the refinement phase to ensure the generation of high-quality, complex code.
\vspace{-2mm}
\subsection{Analysis of Different Base Models Performance}

\reftab{tab:base_model_ds1000} shows the performance results of CoCoST on the DS-1000 dataset with different base models. We can see that GPT-4 has been comprehensively improved with CoCoST, but the performance on GPT-3.5 and WizardCoder is mixed. This indicates that CoCoST requires the model to have the following capabilities to enhance its performance:

\textbf{For code generation planning ability, the higher the complexity of the code that needs to be generated, the higher the demand for planning ability.} Planning capability is key to online retrieval; only correct planning can generate appropriate queries to retrieve useful information. After incorporating online retrieval, GPT-3.5 has an increase of 4.75\% on ClassEval, yet it decreased by 9.2\% on DS-1000 as shown in \reftab{tab:base_model_ds1000}. 
The challenge of ClassEval lies in how to generate the entire class and the interrelated functions, but the complexity of individual function codes is not as high as DS-1000. Thus, the planning ability of GPT-3.5 can handle ClassEval, but it is inferior on DS-1000.

\textbf{Code generation necessitates models to have in-context learning abilities.} The generated code should be built on all the above-provided contents, and the understanding of the preceding input prompt is of great importance in the refinement stage. In \reftab{tab:base_model_ds1000}, it is observed that WizardCoder has a noticeable drop of 18.9\% and 1.1\% on the DS-1000 dataset when utilizing online retrieval and refinement respectively. And the overall performance of WizardCoder is comparatively interior to GPT models. This could be due to WizardCoder's limited in-context learning ability, especially with the complex and lengthy prompts, hindering accurate context comprehension and code modification.

\subsection{Cascade Analysis}
Our framework consists of multiple components cascaded together, which results in certain intermediate steps that cannot be explicitly validated for effectiveness, as well as the potential generation of cascading errors. For the former, a discussion is provided in \refsec{app:pipeline}, while for the latter, an error analysis is conducted in \refsec{app:error}. 

\subsubsection{Analysis of Pipeline}
\label{app:pipeline}
Regarding the generation of test cases, to prove that the generated test cases are comparable to the ground truth test cases, we substitute the generated test cases in CoCoST with ground truth test cases and conduct experiment on the DS-1000 dataset using GPT-4. The results showed that the performance was 68.70\%, only marginally higher by 0.7\% compared to the use of generated test cases (68.00\%). This proves that the effects of both are comparable and that the generated test cases rarely lead to errors.

Regarding generating plans, we attempt to bypass the planning step and directly perform online retrieval. On the DS-1000 dataset using GPT-4, this approach results in a performance of 55.70\%, which is 6.4\% lower than using planning for online retrieval (62.10\%), and even 4.5\% lower than only using GPT-4 to generate code (60.20\%). This confirms that the generated plans are significantly effective.

\begin{figure}[htbp]
    \centering
    \includegraphics[width=1\linewidth]{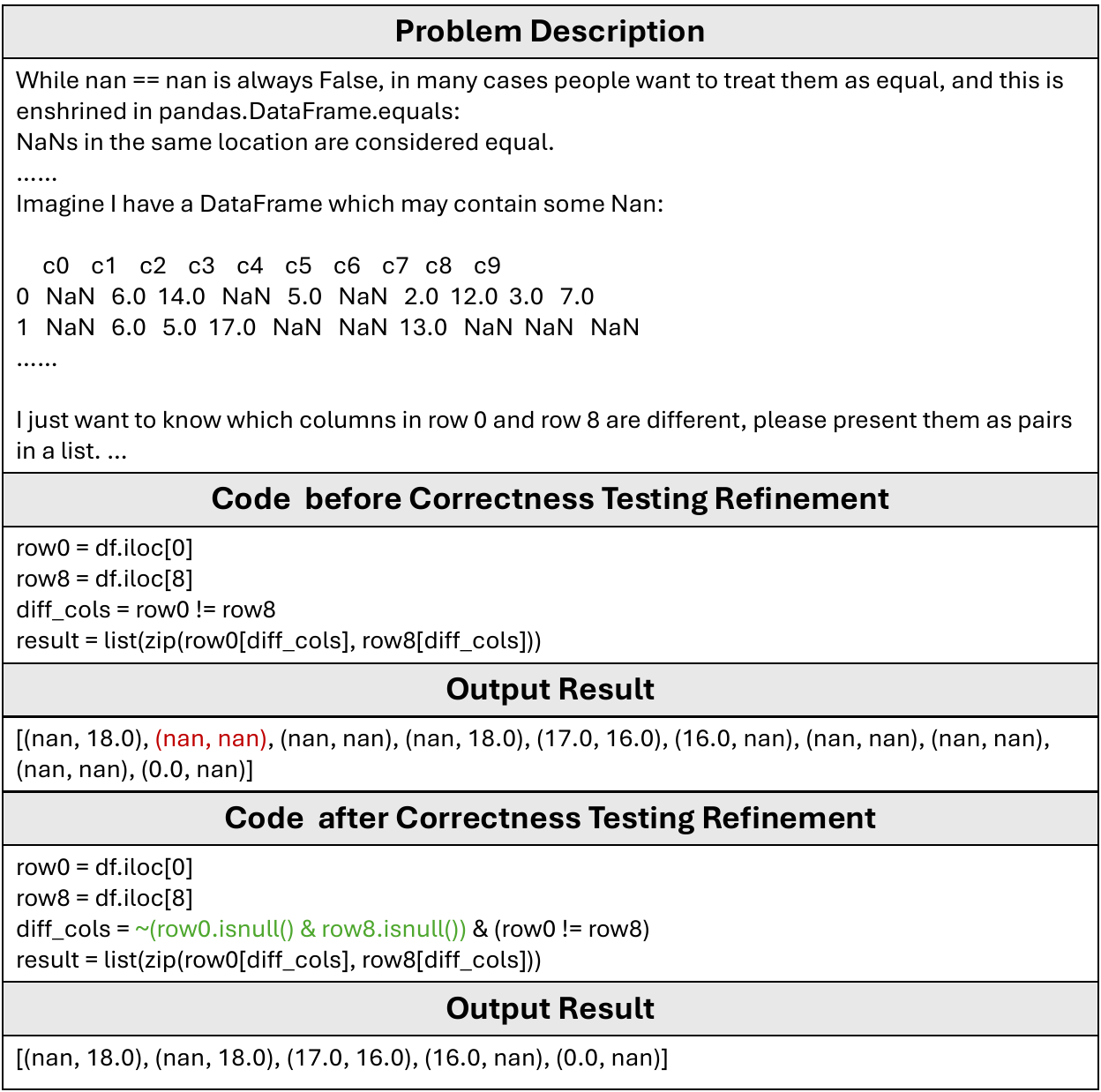}
    \caption{Case Study for Correctness Testing.}
    \label{fig:case_refine}
    \vspace{-3mm}
\end{figure}

\subsubsection{Error Analysis}
\label{app:error}
For cascading errors, some errors generated by retrieval processes can be corrected through refinement, while others may persist, necessitating future improvements. 
Take DS-1000 as an example: Compared to the baseline (GPT-4), we observed that among the instances that turned erroneous after retrieval, 39.6\% were corrected, while 60.4\% remained incorrect. 
For the former cases, it shows that CoCoST can fix some bad cases in the refinement stage even though the retrieval contents have some errors. 
For the later cases, we do observe some cases that are worthy of further research as potential directions for future work. First, the search content could be more detailed. E.g., Some basic steps that LLMs consider unnecessary to search for are not generating queries, but are exactly where the bug is in the code. Second, the search query could be more targeted. The descriptions of some queries are not specific enough in terms of some complicated problems. Therefore, more sub-queries are needed to help the model receive clearer instructions. 

\begin{figure}[htbp]
    \centering
    \includegraphics[width=1\linewidth]{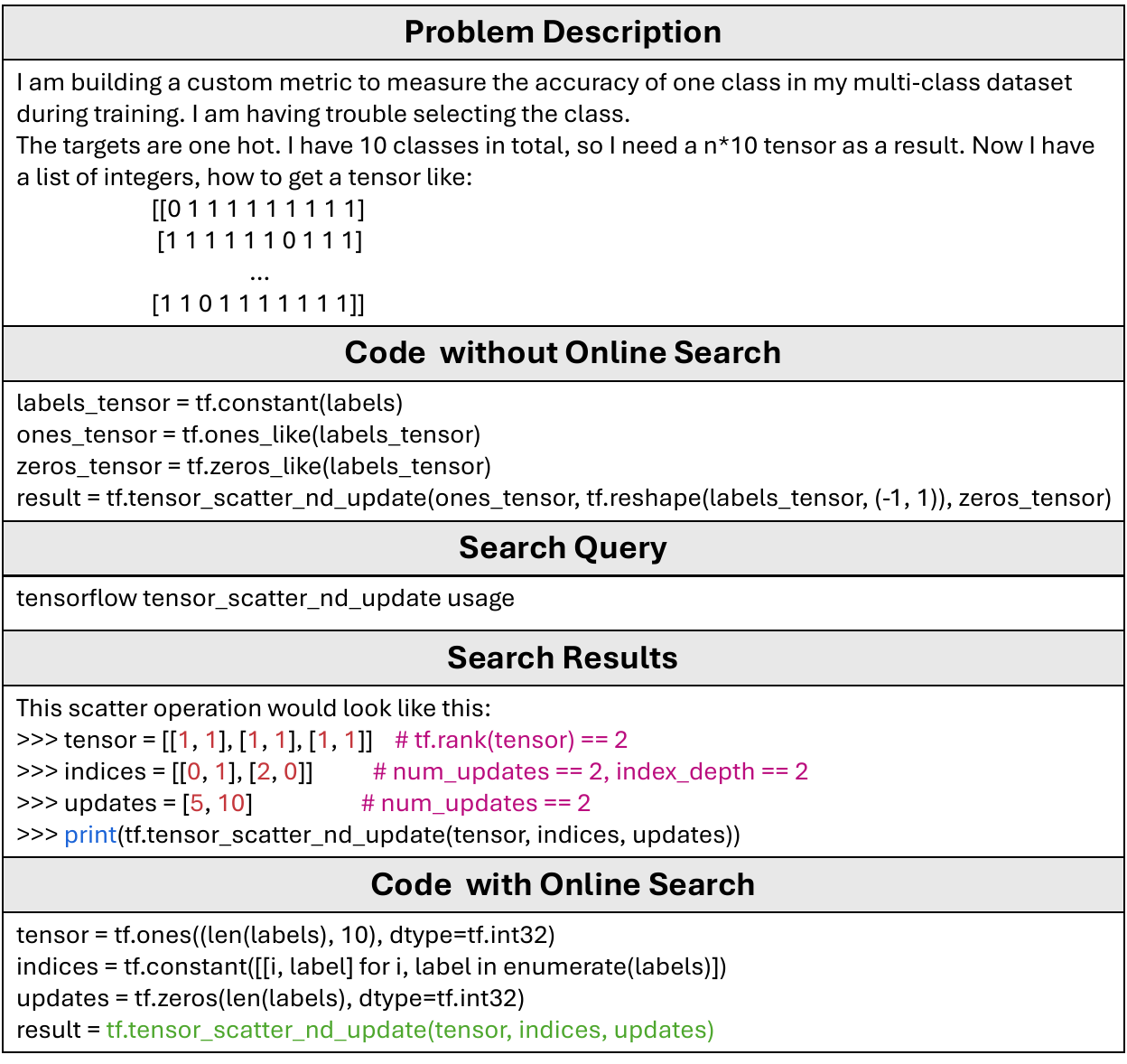}
    \caption{Case Study for Online Retrieval.}
    \label{fig:case_retrieve}
    \vspace{-3mm}
\end{figure}

\subsection{Case Study}

For the case study on online retrieval, refer to \reffig{fig:case_retrieve}. It can be observed that by imitating the usage of functions found through online search, the model is better equipped to prepare the required parameters for the functions and to generate corresponding code. This significantly reduces the difficulty of generating complex code.
For the case study on correctness testing, refer to \reffig{fig:case_refine}. It is evident that, although the initially generated code did not show obvious errors, the output of the code did not align with the expected results. The model refines the code based on the output, thus improving hidden errors and generating the correct code.



\section{Conclusion}
\vspace{-1mm}
In this paper, we propose CoCoST, a novel framework for generating complex code in real-world scenarios by emulating human coding processes like online searching and test case creation. It effectively overcomes challenges in code structure and logic, subtle bug detection, and handling of complex data. The framework's innovative use of online search, planning for query generation, correctness testing, and input-output serialization significantly improves code accuracy and model understanding. Tested on various datasets, CoCoST outperforms existing methods, demonstrating its efficacy in real-world code generation tasks.

\section*{Limitations}
The primary limitation of our research is exceeding the allowed access limits due to frequent calls to the Google Search API. Likewise, we made multiple API calls to test and optimize the performance of the GPT model.

\section*{Ethics Statement}
The datasets and other associated resources utilized in this study are publicly available and widely used in various other existing work.
All the datasets used in this paper have been reviewed to ensure that they do not contain personally identifiable information or offensive content. 
However, since these datasets are sourced from the Internet, potential bias may still be present. 
Furthermore, despite our careful review, the process involving the LLMs may inadvertently introduce inappropriate information into the evolved data.

\section*{Acknowledgments}
We thank all anonymous reviewers for their valuable comments. Xinyi He and Zejian Yuan were supported in part by the National Key R\&D Program of China (2023YFB4704900) and NSFC (61976170, 62088102).


\bibliography{main}

\clearpage
\appendix
\section{Online Searching Detail}
\label{app:search}
Website Base Station Distributions Table During the DS-1000 Online Retrieval Process:
\begin{table}[htbp]
  \centering
  \caption{Website Base Station Distributions Table During the DS-1000 Online Retrieval Process.}
    \begin{tabular}{lc}
    \toprule
    \multicolumn{1}{c}{Website} & \multicolumn{1}{l}{Proportion} \\
    \midrule
    https://stackoverflow.com & 57.92\% \\
    https://numpy.org & 8.59\% \\
    https://pandas.pydata.org & 5.70\% \\
    https://www.geeksforgeeks.org & 5.07\% \\
    https://docs.scipy.org & 4.76\% \\
    https://matplotlib.org & 3.20\% \\
    https://www.tensorflow.org & 3.04\% \\
    http://scikit-learn.org & 2.42\% \\
    The Others & 9.29\% \\
    \bottomrule
    \end{tabular}%
  \label{tab:website}%
\end{table}%

Moreover, CoCoST can be applied to specialized, proprietary, or domain-specific knowledge repositories as long as they are accessible via query. Moreover, implementing queries for private datasets is easily achievable and a growing trend in data management. Major companies such as Google and Microsoft already offer products designed to search private data; for example, Google Workspace's Cloud Search provides powerful capabilities for enterprises to search their private data. In this paper, to validate the effectiveness of our framework, we conducted tests on public online searches. Moving forward, the framework can be applied to an even broader range of knowledge repositories.
\section{Experiment}
\subsection{Datasets Detail}
\label{app:ds-1000}
Further details of DS-1000 implementation are as follows:
\begin{itemize}[leftmargin=0pt,itemindent=\parindent, itemsep=2pt,topsep=0pt,parsep=0pt]
    \item The dataset provides both Insertion and Completion style prompts, where the data is the same, differing only in prompt format, thus yielding similar results. In this paper, experiments are conducted with the Completion style prompt.
    \item We implement a filtering approach to prevent data leakage and model replication of existing solutions from Stack Overflow. The DS-1000 dataset originates from Stack Overflow, and concurrently, over 50\% of the websites we encountered during our online searches are from Stack Overflow. Thus, to prevent data leakage, when conducting online searching, we filter out all Stack Overflow problems belonging to the source of the DS-1000 dataset by using the Stack Overflow question\_id.
\end{itemize}

\subsection{Base Models}
The parameter details for each model in the experiment are as follows:
\begin{itemize}[leftmargin=0pt,itemindent=\parindent, itemsep=2pt,topsep=0pt,parsep=0pt]
    \item GPT-4: model: \textit{gpt-4-32k-0613}, temperature: 0, top\_p: 0.95, max\_tokens: 1024.
    \item GPT-3.5: model: \textit{gpt-35-turbo-16k-0613}, temperature: 0, top\_p: 0.95, max\_tokens: 1024.
    \item WizardCoder: \textit{WizardCoder-Python-13B-V1.0}, temperature: 0, top\_p: 0.95, max\_tokens: 1024.
    
\end{itemize}

\subsection{Baselines Details}
\label{app:baselines}
\begin{itemize}[leftmargin=0pt,itemindent=\parindent, itemsep=2pt,topsep=0pt,parsep=0pt]
    \item DocPrompt~\cite{zhou23docprompting}: DocPrompting enhances the LLM by employing a fine-tuned retriever to fetch problem-relevant documentation from offline document pools. The model then conditions on these documents, along with the problem description, to generate code.
    \item Self-Debugging~\cite{chen2023teaching}: This approach depends on a SQL application or Python interpreter to instruct language models in revising SQL commands or Python code containing errors. For the sake of a fair comparison, we utilize its "simple" variant.
    \item SELFEVOLVE~\cite{jiang2023selfevolve}: Employs LLMs as both sources of knowledge and self-reflective programmers. During the self-reflective process, it refines the code by addressing bugs.
    \item Reflexion~\cite{shinn2023reflexion}: Reflexion utilize reflective feedback with generated tests and episodic memory to process task feedback. For the sake of a fair comparison, we utilize GPT-4 as base model and set trail number = 1.
\end{itemize}

It is worth noting that the test cases involving the refinement process in the baselines mentioned above all use the test cases from the dataset designated for testing. However, within the context of the real-world scenario of CoCoST, test cases from the dataset should not be used within the framework. Without these test cases, they are entirely incapable of functioning. 

\begin{table*}[htbp]
  \centering
  \caption{Table of Main Results for different packages in DS-1000. All metric numbers are represented as percentages. The \textbf{bold} number indicates the highest performance.}
  \resizebox{0.9\textwidth}{!}{
    \begin{tabular}{l|ccccccc|c}
    \toprule
    \multicolumn{1}{c|}{Method} & Pandas & Numpy & Matplotlib & Tensorflow & Scipy & Sklearn & Pytorch & Total/Avg. \\
    \midrule
    CoCoST & \textbf{59.45 } & \textbf{75.91 } & \textbf{75.48 } & \textbf{71.11 } & \textbf{61.32 } & 63.48  & \textbf{77.94 } & \textbf{68.00 } \\
    \ \, + retrieve& 51.89  & 70.91  & 68.39  & 66.67  & 52.83  & \textbf{70.43 } & 60.29  & 62.10  \\
    \ \, + refine& 55.67  & 72.73  & 74.19  & 64.44  & 54.72  & 60.00  & 70.59  & 64.10  \\
    \ \, GPT-4 only& 52.23  & 70.45  & 67.74  & 55.56  & 50.00  & 64.35  & 55.88  & 60.20  \\
    \bottomrule
    \end{tabular}%
    }
  \label{tab:ds-1000_res}%
\end{table*}%

\subsection{Ablation Study Details }
\label{app:abla}
\begin{itemize}[leftmargin=0pt,itemindent=\parindent, itemsep=2pt,topsep=0pt,parsep=0pt]
    \item Without refinement of output: During the refinement process, the output result is not refined; that is, refinement is conducted solely based on the error.
    \item Without refinement of error: During the refinement process, the error is not refined; that is, refinement is conducted solely based on the output result.
    \item Without serialization: During the refinement process, the input and output are not serialized; instead, their printout results are directly used as input.
    \item Without generation of test cases: Test cases are not generated. Since refinement cannot be performed without test cases, only online retrieval is conducted.
    \item Without online retrieval: Online retrieval is not performed, and the process is limited to refinement with correctness testing.

\section{Experimantal Results}

 The main results for different packages in DS-1000 are shown in \reftab{tab:ds-1000_res}.

 The results indicate that CoCoST shows a more pronounced effect on libraries whose inputs and outputs are more complex or more challenging for LLMs to intuitively understand, such as Matplotlib, TensorFlow, and PyTorch. On Sklearn, CoCoST experiences a slight decline due to its test cases containing complex objects, which present a significant challenge in generating test cases. Consequently, CoCoST's performance on Sklearn is not as strong as with the other libraries.

\end{itemize}


\newpage
\section{Prompts of CoCoST}
\begin{figure*}[htbp]
    \centering
    \includegraphics[width=1\linewidth]{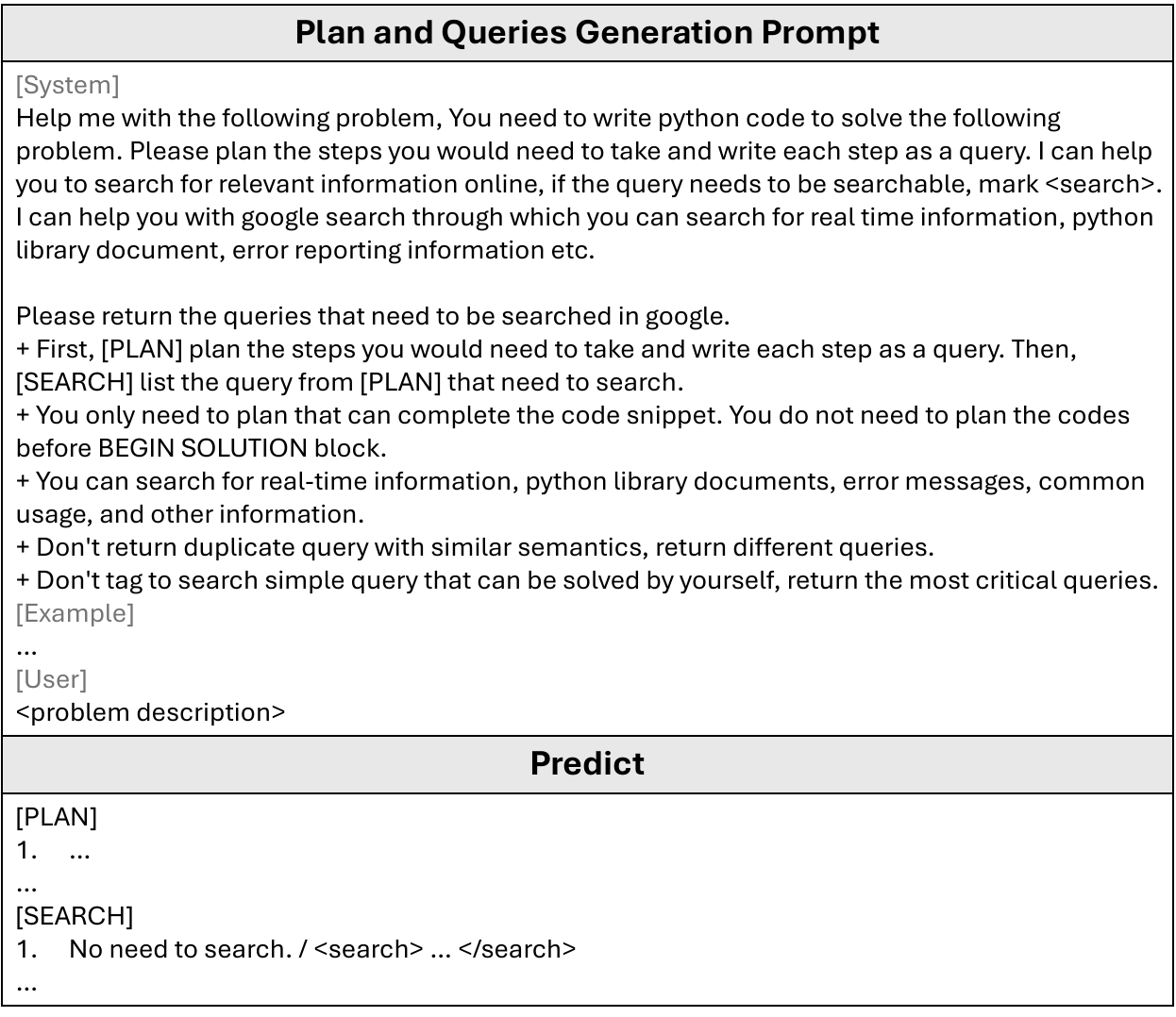}
    \caption{Plan and Queries Generation Prompt on DS-1000.}
\end{figure*}

\begin{figure*}[htbp]
    \centering
    \includegraphics[width=1\linewidth]{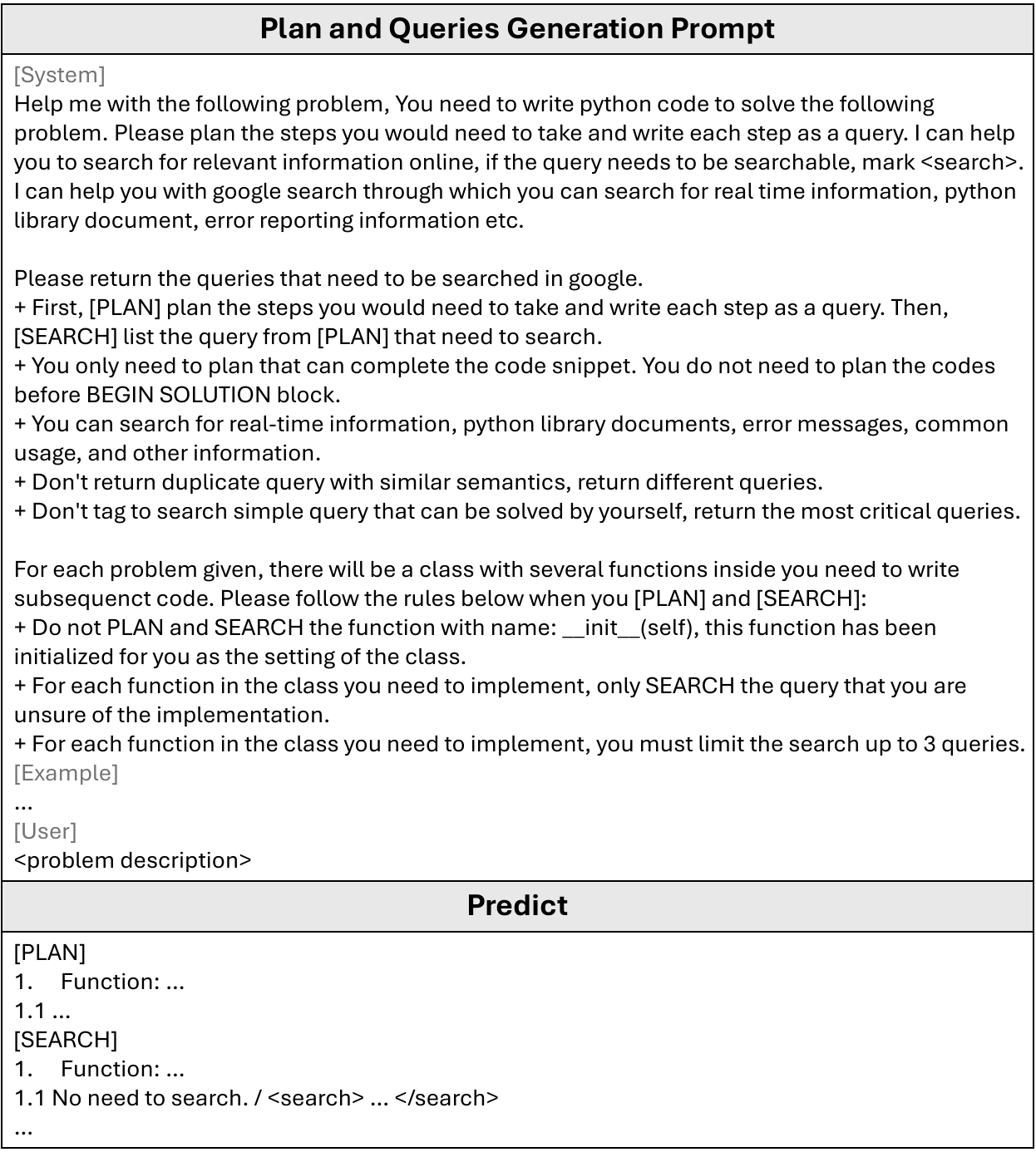}
    \caption{Plan and Queries Generation Prompt on ClassEval.}
\end{figure*}

\begin{figure*}[htbp]
    \centering
    \includegraphics[width=1\linewidth]{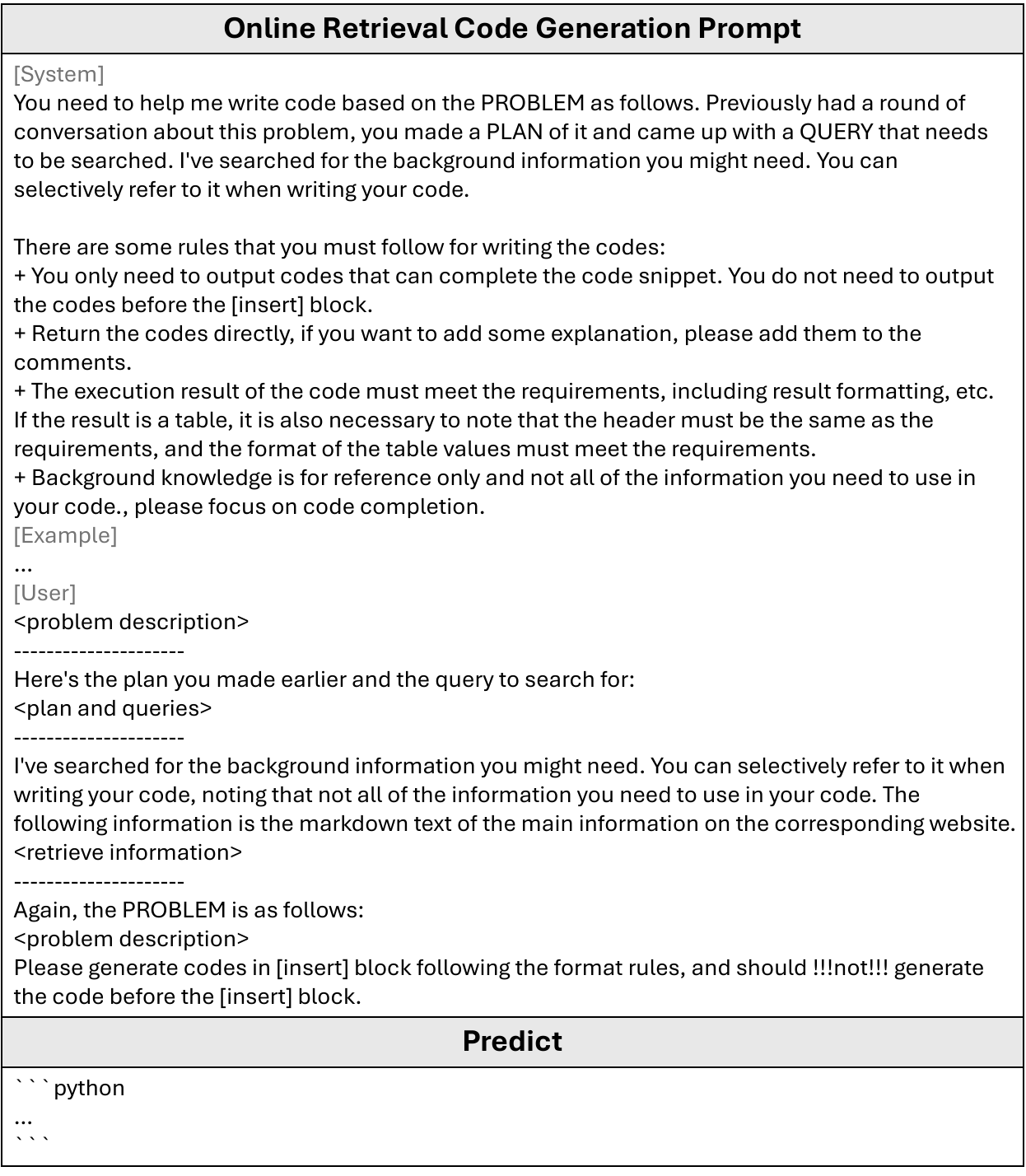}
    \caption{Online Retrieval Code Generation Prompt on DS-1000 and ClassEval.}
\end{figure*}

\begin{figure*}[htbp]
    \centering
    \includegraphics[width=1\linewidth]{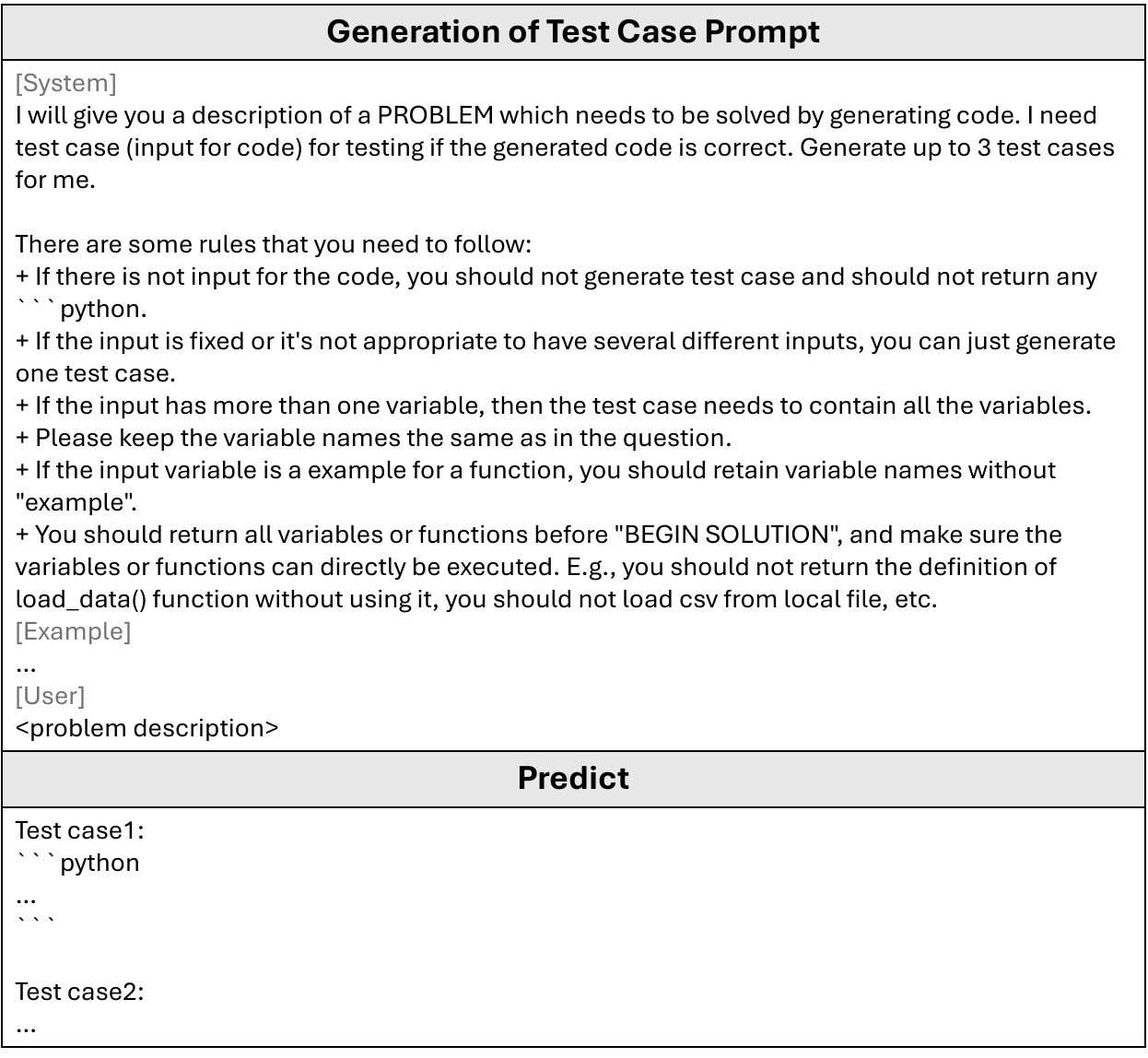}
    \caption{Generation of Test Case Prompt on DS-1000.}
\end{figure*}

\begin{figure*}[htbp]
    \centering
    \includegraphics[width=1\linewidth]{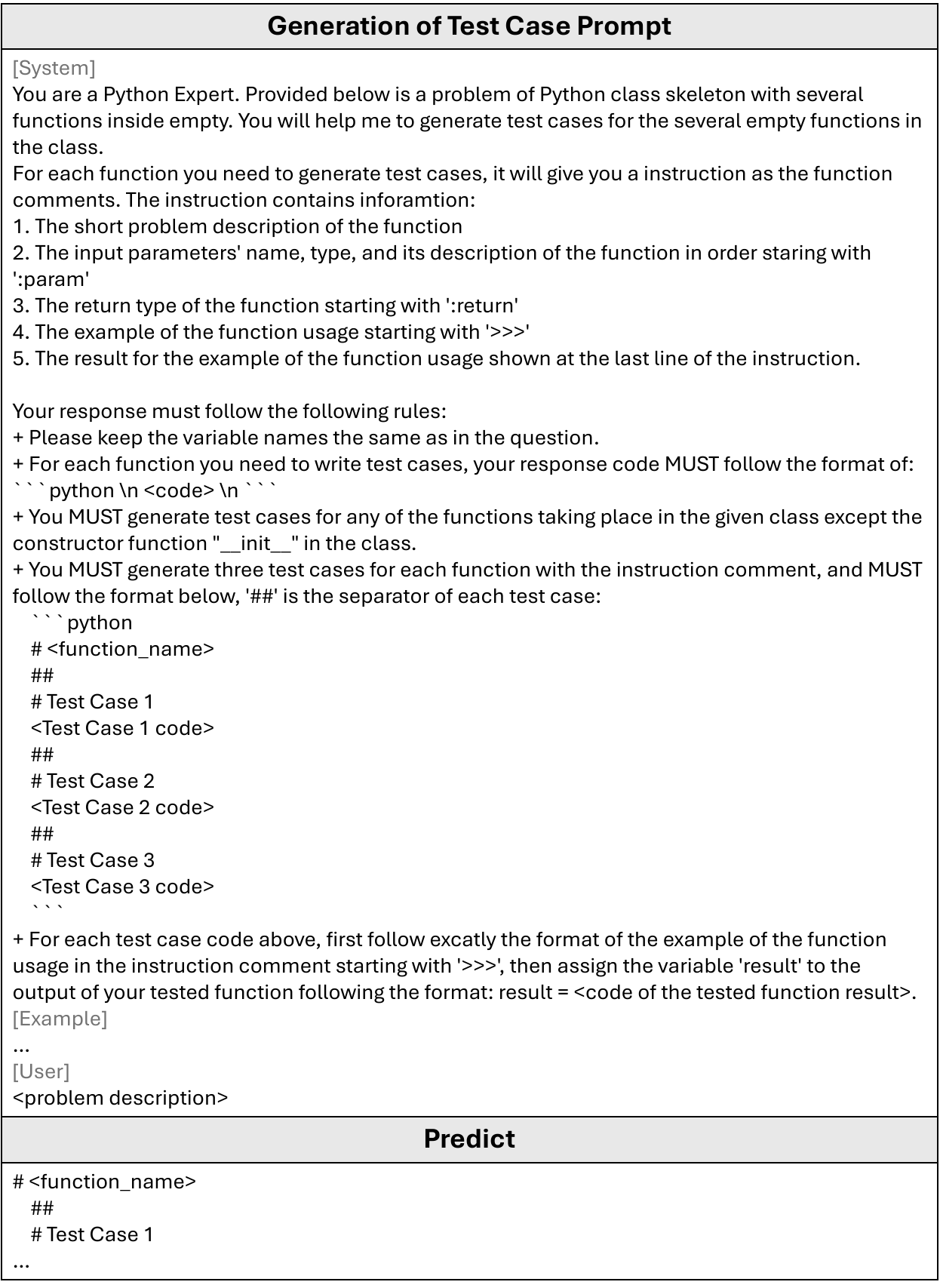}
    \caption{Generation of Test Case Prompt on ClassEval.}
\end{figure*}

\begin{figure*}[htbp]
    \centering
    \includegraphics[width=1\linewidth]{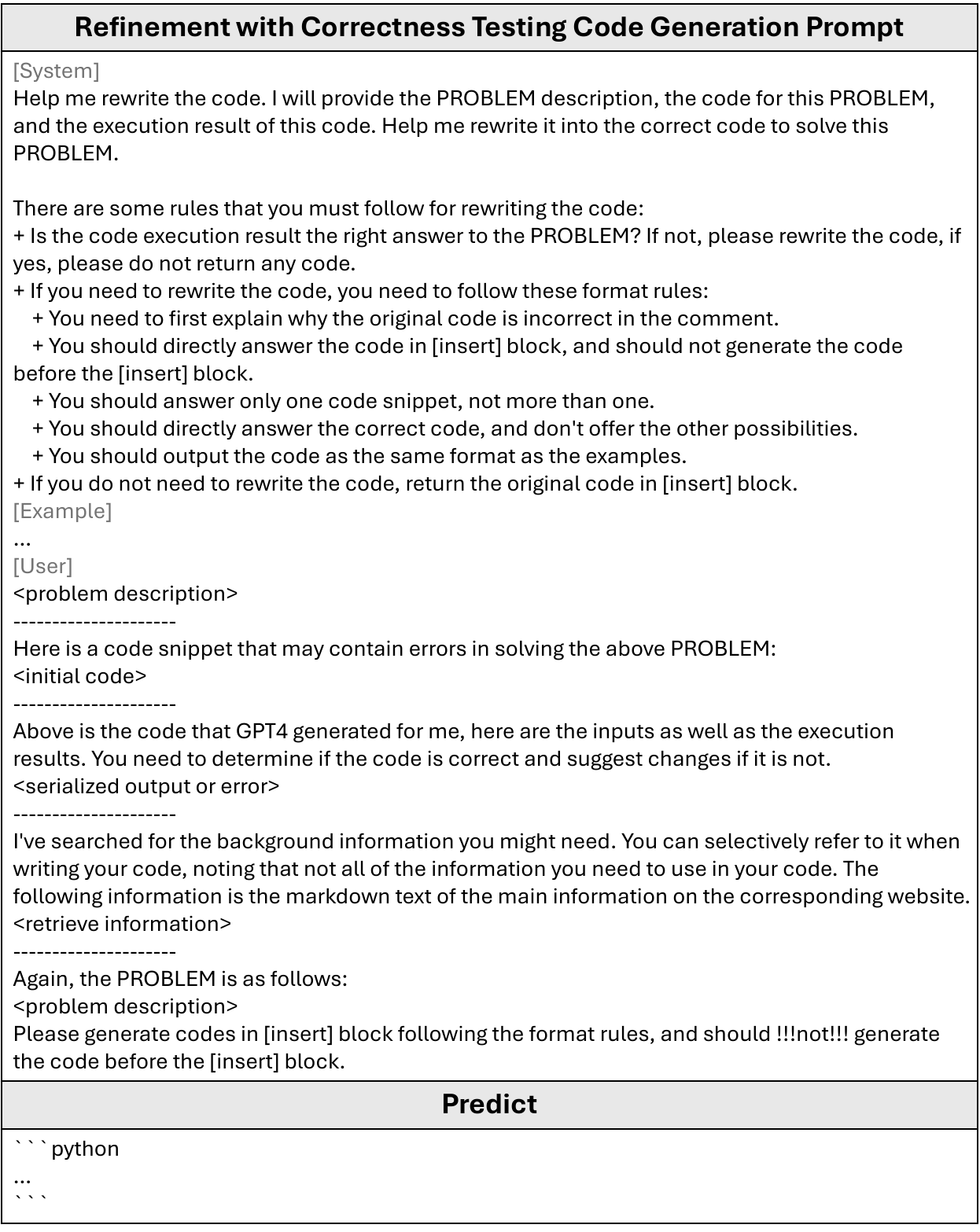}
    \caption{Refinement with Correctness Testing Code Generation Prompt on DS-1000 and ClassEval.}
\end{figure*}

\label{sec:appendix}
\end{document}